\documentstyle[multicol,aps,epsf]{revtex}
\def\part#1#2{\frac{\partial#1}{\partial#2}}
\newcommand\sep{;}
\begin{document}
\title{Scaling Theory for Migration-Driven Aggregate Growth}
\author{F. Leyvraz}
\address{Centro de Ciencias F\'\i sicas, Av.~Universidad
s/n, Col.~Chamilpa, Cuernavaca, Morelos, MEXICO}
\author{S. Redner}
\address{Center for BioDynamics, Center for Polymer Studies, and Department
  of Physics, Boston University, Boston MA 02215}
\date{\today}
\maketitle
\begin{abstract}
  
  We give a comprehensive description for the irreversible growth of
  aggregates by migration from small to large aggregates.  For a homogeneous
  rate $K(i;j)$ at which monomers migrate from aggregates of size $i$ to
  those of size $j$, that is, $K(ai;aj)\sim a^\lambda K(i;j)$, the mean
  aggregate size grows with time as $t^{1/(2-\lambda)}$ for $\lambda<2$.  The
  aggregate size distribution exhibits distinct regimes of behavior that are
  controlled by the scaling properties of the migration rate from the
  smallest to the largest aggregates.  Our theory applies to diverse
  phenomena such as the distribution of city populations, late stage
  coarsening of non-symmetric binary systems, and models for wealth exchange.

\end{abstract}
\pacs{82.20.-w, 05.40.-a, 89.65.Cd, 89.75.Fb}
\begin{multicols}{2}

Much attention has been devoted to understanding the irreversible growth of
aggregates through binary coalescence.  This general mechanism arises in
diverse branches of physics, such as gelation\cite{gel}, island formation in
epitaxial surface growth\cite{surf}, and stellar evolution\cite{astro}.  By a
long-term research effort, considerable understanding of this irreversible
aggregation process has been achieved\cite{Ernst,scaling}.  In this Letter,
we focus on a different growth mechanism that appears to provide a natural
description for the evolution of city populations.  This is preferential
evaporation from smaller aggregates and preferential condensation onto larger
aggregates
  
There are many examples where this evaporation/condensation mechanism occurs
in physics and in the social sciences.  The classic physics example is the
late-stage coarsening of a binary mixture in an off-critical quench below but
near the coexistence curve\cite{LSW,Bray}.  Here the system separates into
droplets of the minority phase that are embedded in a matrix of the majority
phase.  Subsequent growth proceeds through preferential evaporation from
smaller droplets, due to the effect of surface tension, and subsequent
condensation onto the larger droplets\cite{LSW}.

In the social sciences, it has been argued \cite{Zhang} that the growth of
cities may be due to migration from small to large cities, as opposed to a
view that emphasizes differential population growth \cite{Zanette}.  In a
spirit closer to our work, the wealth distribution of individuals was
described by a kinetic asset exchange model with preferential transfer from
poor to rich individuals in each transaction\cite{Redner}.  For generic
situations, the solution to the rate equation showed that this mechanism
gives very different kinetic behavior from conventional aggregation.

Motivated by these fragmentary results, we investigate a general class of
migration-driven growth phenomena and show that, at large times and for large
aggregate sizes, a comprehensive scaling theory can be developed with a
minimum of assumptions.  This theory gives both the growth rate of the
typical aggregate size, as well as the asymptotics of the aggregate size
distribution.  Basic features of our theory agree with data on the population
dynamics of U.S. cities.  An important feature of our theory is that one may
infer the general form of the migration rates from observations of the
aggregate size distribution.  For phenomena such as the city population
distribution or the wealth distribution, we may thus hope to predict basic
aspects of the dynamics in systems for which we have little {\it a priori\/}
knowledge of underlying microscopic driving mechanisms.

The model that we study is defined as follows.  Let aggregates $A_j$ be
characterized only by their mass $j$, or equivalently, by the number of
individuals that comprise them.  These aggregates evolve according to the
following irreversible reaction
\begin{equation}
\label{RE}
A_k+A_l\mathop{\longrightarrow}^{K(k\sep l)}A_{k-1}+A_{l+1}\qquad{k\leq l}.
\end{equation}
That is, a monomer (or equivalently, one person) leaves a smaller aggregate
of size $k$ and joins a larger one of size $l$ with rate $K(k;l)$.  This
generalizes the asset exchange model of Ref.~\cite{Redner}, where a
restricted class of reaction rates $K(k;l)$ were considered.  More generally,
migration could also go from a larger to a smaller aggregate.  The symmetric
limit, where the migration direction does not depend on the relative sizes of
the two aggregates, leads to a diffusive-like kinetic universality class.  We
defer the investigation of this general system to a future work.  Instead, we
focus on the situation where there is preferential migration from small to
large aggregates.  In fact, {\em any} migration bias leads to scaling
behavior for the aggregate size distribution identical to that of complete
bias, as embodied by Eq.~(\ref{RE}).

We now make the assumption of spatial homogeneity, so that the system is
fully characterized by the concentrations $c_j(t)$ of aggregates $A_j$ of
size $j$ at time $t$.  We also assume that the law of mass-action applies so
that the time dependence of the aggregate concentrations may be described by
the following rate equations:
\begin{equation}
\label{eq:smol}
\dot c_j(t)=\frac{1}{2}\sum_{k,l=1}^\infty K(k\sep l)c_k(t)c_l(t)\left[
\delta_{k,j+1}+\delta_{l,j-1}-\delta_{k,j}-\delta_{l,j}
\right].
\end{equation}
The various delta-function terms enforce the constraint that the size of each
aggregate changes by $\pm 1$ in a single reaction.  The initial condition may
be taken to be $c_j(0)=\delta_{j,1}$, but any initial condition may equally
well be considered, provided it is rapidly decaying in $j$.  

{}From these equations, we can immediately draw several important
conclusions.  First, there are no equilibrium solutions.  Rather, the size of
aggregates grows continuously and each $c_j(t)$ eventually goes to zero as
$t\to\infty$.  Second, the total mass contained in the aggregates is
(formally, at least) conserved.  That is,
\begin{equation}
\label{eq:mass}
\frac{d}{dt}\sum_{j=1}^\infty jc_j(t)=0,
\end{equation}
if the necessary interchanges between the infinite sums in this equation can
be justified.  Here we shall confine ourselves to this mass-conserving case.
For definiteness we normalize the total mass to unity.

We now make the conventional scaling ansatz for the large-time behavior of
$c_j(t)$\cite{scaling}.  We assume that there exists a well-defined typical
aggregate size $s(t)$ at time $t$ such that
\begin{equation}
\label{eq:scaling}
c_j(t)\to j^{-2}\Phi\big(j/s(t)\big).
\end{equation}
Here the exponent $-2$ follows directly from the condition that the total
mass is conserved, as discussed, for example, in \cite{Ernst}.
We further assume that the reaction rates $K(k\sep l)$ are {homogeneous\/} of
degree $\lambda$, or at least, that they are asymptotically so in the limit
of large sizes.  That is
\begin{equation}
\label{eq:deflambda}
K(ak\sep al)=a^\lambda K(k\sep l)[1+o(1)].
\end{equation}

In the context of city population growth, the homogeneity exponent $\lambda$
can be given the following interpretation.  When the populations of two
cities are scaled by some factor, there are both more susceptible migrants in
the smaller city and potentially more reasons to move to the larger city.  It
is then natural that the overall migration rate varies as a power law in this
scale factor.  Exceptions to this behavior typically involve the existence of
a cutoff size that separates two qualitatively different kinds of behaviors.

Substituting the scaling ansatz (\ref{eq:scaling}) into Eq.~(\ref{eq:smol}),
we find that $s(t)$ satisfies $\dot s(t)=s(t)^{\lambda-1}$, with asymptotic
solution, for $\lambda<2$,
\begin{equation}
\label{eq:defsize}
s(t)\propto\left[(2-\lambda)t\right]^{1/(2-\lambda)}.
\end{equation}
Defining $z$ as the growth exponent of $s(t)$, we thus have
\begin{equation}
\label{DefZ}
z=\frac{1}{2-\lambda}.
\end{equation}
This growth exponent can also be obtained by adaptation of a
back-of-the-envelope estimate for the typical size in irreversible
aggregation.  In aggregation, the reaction of aggregates of typical size $s$
leads to a growth $\Delta s$ of the order of $s$ in a time $\Delta t$ of the
order of $1/({\rm concentration}\times s^{-\lambda})$.  Here the
concentration scales as $1/s$ and $s^{-\lambda}$ is the inverse reaction rate
between typical-size aggregates.  This leads to $\dot s\sim s^\lambda$, from
which $s\sim t^{1/(1-\lambda)}$.  For migration-driven growth, $\Delta s$ is
now of the order of 1 in the time $\Delta t$.  This gives $\dot s\sim
s^{\lambda-1}$, thus reproducing the growth exponent of Eq.~(\ref{DefZ}).

On the other hand, if $\lambda>2$, a power-law decay of the $c_j(t)$ in $j$
sets in at finite time.  This feature invalidates the mass conservation
statement and hence the scaling form of Eq.~(\ref{eq:scaling}).  The limiting
case $\lambda=2$ can be treated within our scaling formulation, but must be
handled with particular care, as we discuss below.  This pattern of behavior
for the time dependence of the typical size parallels that of conventional
aggregation, except that the size exponent in aggregation is $z=
1/(1-\lambda)$ and a finite-time gelation transition occurs for $\lambda>1$
\cite{gel,Ernst}.  Note also that when the migration rate is symmetric, a
scaling analysis similar in spirit to that just presented shows that the mean
aggregate size grows as $t^{1/(3-\lambda)}$ for $\lambda <3$.  Thus even
migration without population bias leads to growing aggregates, albeit at a
slower rate than if a bias towards larger aggregates exists.

Also from the scaling ansatz, we find, after some non-trivial algebra, that
the scaled aggregate size distribution $\Phi(x)$ obeys
\begin{mathletters}
\begin{equation}
\label{eq:DiffPhi}
\frac{d\Phi(x)}{dx}=-x\frac{d}{dx} \left[\frac{\Phi(x)\Psi(x)}{x^2}\right],
\end{equation}
with 
\begin{equation}
\label{eq:DefPsi}
\Psi(x)=\int_0^\infty\frac{dy}{y^2}\left[K(x\sep y)-K(y\sep x)
\right]\Phi(y).
\end{equation}
\end{mathletters}
From these equations the basic qualitative behavior of $\Phi(x)$ can be
deduced.  Note that only the antisymmetric part of $K(x\sep y)$ contributes
to the scaling limit.  Without loss of generality, we can now assume that
$K(k\sep l)=0$ for $k>l$.  From Eq.~(\ref{eq:DiffPhi}) it follows that
$\Phi(x)$ can be discontinuous whenever $\Psi(x)+x$ becomes zero. In
particular, at such a point $x_c$, $\Phi(x)$ can be consistently set to zero
for all $x>x_c$. As long as $\Phi(x)$ is different from zero, however,
(\ref{eq:DiffPhi}) can be integrated to yield
\begin{equation}
\label{eq:IntPhi}
\Phi(x)=\frac{Ax^2}{x+\Psi(x)}\exp\left[-\int_{x_0}^x
\frac{dy}{y+\Psi(y)}\right].
\end{equation}
Here $A$ and $x_0$ are arbitrary constants chosen so that
\begin{equation}
\label{eq:MassNorm}
\int_0^\infty\frac{\Phi(y)}{y}dy=\sum_{j=1}^\infty jc_j(0)=1.
\end{equation}

An important feature of $\Phi(x)$ is its behavior for small values of $x$.
To quantify this, we define the exponent $\tau$ through
\begin{equation}
\label{eq:DefTau}
\Phi(x)\propto x^{2-\tau}\left[1+o(1)\right].
\end{equation}
With this definition, one has $c_j(t)\sim j^{-\tau}$ for $1\ll j\ll s(t)$, as
well as
\begin{equation}
\label{eq:DefW}
c_j(t)\sim t^{-(2-\tau)z}\equiv t^{-w}\qquad(1\ll j\ll s(t)).
\end{equation}
This defines the exponent $w$. 

To proceed further, we introduce another fundamental exponent that completes
the scaling characterization of the reaction rates $K(k\sep l)$, namely,
\begin{equation}
\label{eq:defmu}
K(1\sep l)\approx l^{\lambda-\mu}\qquad(l\to\infty).
\end{equation}
This is entirely analogous to the corresponding definition in conventional
aggregation where the form of the cluster size distribution depends on the
relative rates of small-small, large-large, and large-small
reactions\cite{Ernst}.  With these definitions, we find, after detailed
analysis of Eq.~(\ref{eq:IntPhi}), four different classes of
behavior:
\begin{itemize}
  
\item {\bf Type 1:} $\lambda\geq1$, $\mu>1$.  In this case $\tau=\lambda$ and
  hence $w=1$.
  
\item {\bf Marginal:} $\lambda>1$, $\mu=1$.  Here it is not possible to make
  simple statements about the value of $\tau$. Rather, $\tau$ depends on the
  complete shape of $\Phi(x)$ and therefore on the very specific form of the
  reaction rates.
  
\item {\bf Type 2a:} $\lambda<1$, $\mu\leq(1+\lambda)/2$.  In this case
  $\tau=\mu$ and $w=(2-\mu)/(2-\lambda)$.
  
\item {\bf Type 2b:} $\lambda<1$, $(1+\lambda)/2\leq \mu<1$. In this case
  $\tau= (1+\lambda)/2$ and $w=(3-\lambda)/(4-2\lambda)$.
\end{itemize}

For the complementary large-$x$ behavior of $\Phi(x)$, we now show that in
almost all cases $\Phi(x)$ vanishes beyond a certain critical value $x_c$ of
its argument.  Indeed, suppose the contrary.  It then follows from
Eq.~(\ref{eq:DefPsi}) that $\Psi(x)\to-\infty$ as $x\to\infty$. This can
happen in three ways: either $\Psi(x)$ varies faster than linear, slower than
linear, or linearly in $x$.  In the first case, Eq.~(\ref{eq:IntPhi}) would
indicate that $\Phi(x)<0$ for large $x$, which is impossible.  In the second
case, $\Phi(x)$ would go to a constant as $x\to\infty$, in contradiction to
Eq.~(\ref{eq:MassNorm}).  Thus the only viable possibility is the third case,
which occurs if $\lambda-\mu=1$.

A more thorough investigation is required to determine whether it is possible
to find a consistent large-$x$ behavior for $\Phi(x)$ in this last case.  If
so, then $\Phi(x)$ would have a power-law decay such that the integral in
Eq.~(\ref{eq:MassNorm}) still converges.  However, in all other cases, 
Eq.~(\ref{eq:IntPhi}) must cease to be valid at some point and the function
$\Phi(x)$ must vanish identically afterwards.  This can happen in two
different ways: Either $x+\Psi(x)$ has a simple zero at some point $x_c$ and
the function $\Phi(x)$ jumps from its value at $x_c$ to zero, which is
possible according to Eq.~(\ref{eq:DiffPhi}), or else the function $\Phi(x)$
goes smoothly to zero at $x_c$ as a consequence of Eq.~(\ref{eq:IntPhi}) and
the double zero of $x+\Psi(x)$ at $x_c$.  These results closely correspond to
those of the Lifshitz-Slyozov-Wagner (LSW) theory of coarsening\cite{LSW}, as
well as to models of asset exchange\cite{Redner}.

A special case that can be solved exactly is the case $\mu=\lambda$.  For
this situation, we find
\begin{eqnarray}
\label{eq:ExactPhi}
\Phi(x)&=&x^{2-\lambda}\qquad(x\leq(2-\lambda)^{1/(2-\lambda)})\\
\Phi(x)&=&0\qquad\text{otherwise.}\nonumber
\end{eqnarray}
If $\lambda>1$, this case belongs to a system of Type 1 listed above, whereas
if $\lambda<1$, this case belongs to Type 2a.  For either alternative, the
correct exponent $\tau$ is predicted.  Note further the discontinuity in
$\Phi(x)$ that indeed occurs exactly at the point where $x+\Psi(x)$ vanishes.
Many other cases can be handled similarly and will be presented in a
forthcoming publication \cite{longpaper}.

A situation that requires a more refined analysis is $\lambda=2$ and $\mu>1$.
For these parameter values, it follows that $\tau=2$, which is incompatible
with the normalization condition Eq.~(\ref{eq:MassNorm}).  To obtain valid
results, we need to modify the scaling ansatz as follows
\begin{equation}\label{eq:ModScaling}
c_j(t)\approx\frac{j^{-2}}{\ln s(t)}\Phi\big(j/s(t)\big).
\end{equation}
It follows that
\begin{equation}
\label{eq:ModSize}
s(t)=\exp\left[\sqrt{2(t+B)}\right],
\end{equation}
where $B$ is some constant.  The function $\Phi(x)$ then has the
normalization $\Phi(0)=1$ and satisfies a modified version of
Eq.~(\ref{eq:DiffPhi}).

Let us now discuss how our scaling theory applies to LSW
coarsening\cite{LSW}.  For this system, the migration rate $K(i\sep j)$ is
given by the product of the rate at which a particle evaporates from an
aggregate of mass $i$ and the probability that it reaches an aggregate of
size $j$.  In the evaporation step, the diffusive current $J$ is $\Delta
c/R(i)$, where $\Delta c$ is the difference between the monomer concentration
near the interface and in the bulk.  This difference is proportional to
$R(i)^{-1}$, since it is due to surface tension.  The current $J$ is
therefore of the order of $R(i)^{-2}$ and thus the rate at which particles
leave an aggregate of size $i$ is proportional to $JR(i)^{d-1}$, that is, to
$R(i)^{d-3}$.  Further, the probability of reaching an aggregate of size $j$
in three dimensions is simply proportional to its volume $R(j)^d$.  We
therefore find for the overall migration rate
\begin{equation}
\label{eq:LSW}
K(i\sep j)\approx R(i)^{d-3}R(j)^d\approx i^{(d-3)/d}j.
\end{equation}

From the definitions of $\lambda$ and $\mu$ in Eqs.~(\ref{eq:deflambda}) and
(\ref{eq:defmu}), it follows that the system is of Type 2a, from which one
obtains $z=d/3$ and $\tau=1-3/d$.  These indeed correspond to the LSW
predictions, in which the characteristic cluster radius increases as
$t^{1/3}$\cite{LSW} and the number $n(R)$ of clusters of radius $R\ll R_c(t)$
varies as $R^2$ \cite{size-radius}.  Our theory also correctly predicts that
the scaling function vanishes beyond a certain value of the scaling variable.
On the other hand, the migration rate of Eq.~(\ref{eq:LSW}) is not precise
enough to ensure that our scaling theory reproduces the same functional form
for $\Phi(x)$ as that of the detailed LSW theory.

To apply our theory meaningfully to the evolution of city populations, it is
necessary to incorporate the effects of demographic population growth.  Over
intermediate time scales (of the order of decades), demographic growth
typically gives a population that increases exponentially with time.  Such a
behavior can be modeled by allowing the process
$A_k\mathop{\longrightarrow}A_{k+1}$ to occur at rate ${k\gamma}$.  When
demographic growth and migration occur together, the scaling ansatz for the
underlying rate equations needs to be modified accordingly.  We have found
that the appropriate scaling ansatz for this more realistic situation is
\begin{equation}
\label{eq:GrowScaling}
c_j(t)\approx j^{-2}e^{\gamma t}\,\Phi\big(j/s(t)\big).
\end{equation}
With this hypothesis, the functional form of $\Phi(x)$ turns out to remain
the same as the case of no demographic growth, but now the typical city
population grows as
\begin{equation}
\label{eq:s-growth}
s(t)\sim e^{\gamma t/(2-\lambda)}.  
\end{equation}
Hence we arrive at the central conclusion that the typical city size grows
much faster than the population of the country as a whole as $\lambda$
approaches 2.

The city population distribution in many countries is consistent with a
power-law form in which the exponent $\tau$ is close to
2\cite{Zhang,Zanette,Zipf}.  Our scaling theory then requires that the
homogeneity exponent $\lambda$ is also close to 2.  Thus from
Eq.~(\ref{eq:s-growth}), the typical city population should increase much
faster than the overall population.  This is confirmed qualitatively by data
for the populations of various U.S. cities during their early histories
\cite{census}.  The population of essentially every major U.S. city grows
much faster than the U.S. as a whole over considerable time range.  However,
as cities reach maturity, their growth may slow or their population may even
decline for reasons unrelated to preferential migration to still larger
cities.

In summary, we have introduced a simple kinetic description for
migration-driven growth and developed a scaling theory that determines the
large-time behavior for the aggregate size distribution.  Asymptotic results
depend only on rudimentary properties of the reaction rates, most notably the
homogeneity index $\lambda$.  The typical aggregate size grows as
$t^{1/(2-\lambda)}$, while several distinct behaviors emerge for aggregate
size distribution.  Our results represent the counterpart of the scaling
theory of irreversible aggregation to migration-driven growth.  Finally, we
have suggested a connection between migration-driven growth to the
distribution of city populations and found a qualitative correspondence
between model predictions and recent data on U.S. cities.

We are grateful to grants NSF INT9600232, NSF DMR9978902, and DGAPA IN112200
for financial support of this work.

\end{multicols}

\begin{references}
\bibitem{size-radius} Note that the equivalence between the number 
$n(R)$ of clusters of a given radius and the number $c_j(t)$ of clusters
of a given mass $j$ is given by $c_j(t)\approx R(j)^{-(d-1)}
n[R(j)]$.
\end{references}
\end{document}